\documentclass[aps,prb,twocolumn,notitlepage,superscriptaddress,showpacs]{revtex4-1}
\usepackage{graphicx,subfigure,epsfig}
\usepackage{dcolumn}
\usepackage{amssymb}
\usepackage{times}
\usepackage{amsmath}
\usepackage{amsfonts}
\usepackage{mathrsfs}
\usepackage{setspace}
\usepackage{latexsym}
\usepackage{bbm}
\usepackage{float}
\usepackage{flafter}
\usepackage{bm}
\usepackage{epstopdf}
\usepackage{hyperref}
\usepackage{color}
\usepackage{multirow}

\newcommand{\E}{\mathrm{e}}

\begin{document}

\title{Breakdown of the Bloch-wave behavior for a single hole in a gapped antiferromagnet}
\author{Zheng Zhu}
\affiliation{Institute for Advanced Study, Tsinghua University, Beijing, 100084, China}
\affiliation{Department of Physics, Massachusetts Institute of Technology, Cambridge, MA, 02139, USA}
\author{D. N. Sheng}
\affiliation{Department of Physics and Astronomy, California State University, Northridge, CA, 91330, USA}
\author{Zheng-Yu Weng}
\affiliation{Institute for Advanced Study, Tsinghua University, Beijing, 100084, China}
\affiliation{Collaborative Innovation Center of Quantum Matter, Tsinghua University, Beijing, 100084, China}

\date{\today}


\begin{abstract}
Whether a doped hole propagates as a Bloch wave or not is an important issue of doped Mott physics. Here we examine this problem based on the quasiparticle spectral weight $Z$ distribution, calculated by density matrix renormalization group (DMRG). By tuning the anisotropy of a two-leg $t$-$J$ ladder without closing the background spin gap, the $Z$ distribution unambiguously reveals a transition of the single hole state from a Bloch wave to a novel one with spontaneous translational symmetry breaking. We further establish a direct connection of such a transition with a nonlocal phase string entanglement between the hole and quantum spins,
which explains numerical observations.
\end{abstract}

\maketitle

For a weakly interacting band insulator, a doped charge behaves like a Bloch wave in the presence of a periodic lattice obeying the Bloch theorem. One may also ask a meaningful question concerning the fate of a hole injected into a Mott insulator with correlated quantum spins\cite{Anderson,Lee2006}. For the special case that such a spin system is gapped and translationally invariant, based on the conventional wisdom, the doped hole would be expected to only disturb its surrounding spins to form some sort of spin polaron \cite{Lee2006,SCBA1,SCBA2,SCBA3,SCBA4}. One might be tempted to generally conclude a Bloch-wave behavior for such a hole doping into a gapped
spin system.

However, recent density matrix renormalization group (DMRG) studies \cite{ZZ2013,ZZ2014,ZZ2014qp,ZZ2014cm}
of hole-doped two-leg spin ladders have revealed an unexpected rich phenomenon even if the undoped system remains gapped. Although an injected hole does propagate like a simple Bloch wave in the strong anisotropic limit of the model,
it undergoes a quantum transition to a novel state when the anisotropy is reduced. After the transition, the charge loses its phase coherence over a finite length scale \cite{ZZ2013,ZZ2014qp}%Wang2015
, concomitant with an emergent interference pattern \cite{ZZ2014cm}
%,Wang2015,ZZ2015
(charge modulation) breaking the translational symmetry.  Besides, a strong pairing of two holes also has been found \cite{ZZ2014} in this regime.

Nevertheless, in a recent new DMRG study of the same model, White, Scalapino, and Kivelson (WSK) claimed \cite{WSK2015} that the critical point seen in the above studies\cite{ZZ2013,ZZ2014,ZZ2014qp,ZZ2014cm}  only signals a qualitative change of the quasiparticle energy  spectrum without changing the Bloch wave nature on the both sides. The WSK's conclusion is largely based on the \emph{total} quasiparticle spectral weight $Z_{\mathrm {tot}}$ (to be defined below), which remains finite and smooth across the transition point in their DMRG calculation \cite{WSK2015}.
%-------------------------------------------------------------------------------Fig.1-----------------------------------------------------------------------------------------------------------
\begin{figure}[tbp]
\begin{center}
\includegraphics[width=0.48\textwidth]{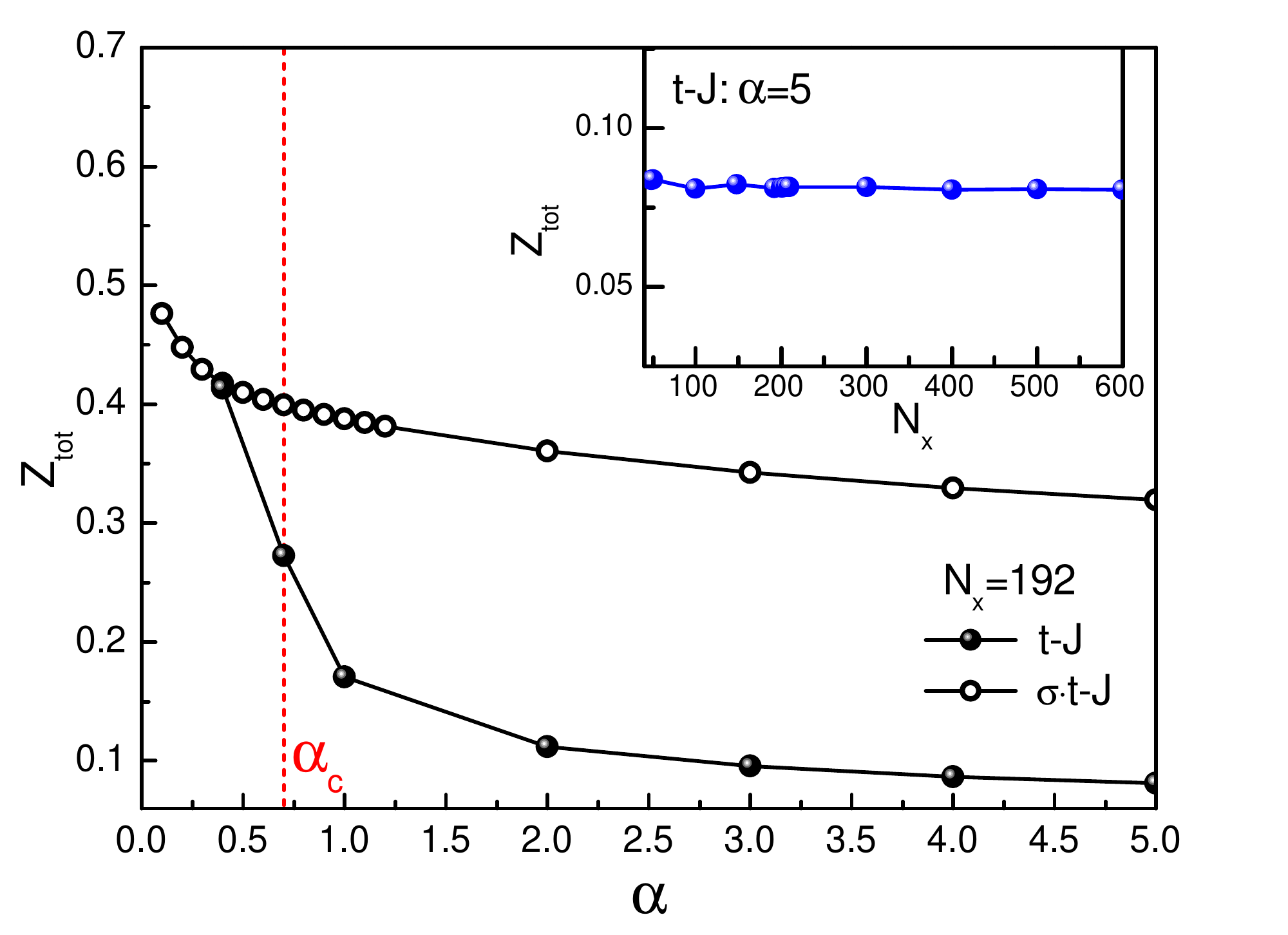}
\end{center}
\par
\renewcommand{\figurename}{Fig.}
\caption{(Color online). $Z_{\mathrm{tot}}$ measures the overlap of the true ground state of a single hole with a bare hole state [cf. Eq.~(\ref{ztot})]. $\alpha$ denotes an anisotropic parameter for the two-leg $t$-$J$ ladder. A critical point $\alpha_c$ is marked by the vertical dashed line, which is previously determined \cite{ZZ2014qp,ZZ2014cm} by DMRG for the $t$-$J$ case at $t/J=3$ with $\alpha_c\approx 0.7$ (but there is no critical $\alpha_c$ for the so-called $\sigma$$\cdot$$t$-$J$ model, see text). Inset: the convergence of $Z_{\mathrm{tot}}$ with the sample size $N= N_x\times 2$ at $\alpha=5$.  }
\label{Fig1}
\end{figure}
%%%%%%%%%%%%%%%%%%%%%%%%%%%%%%%%%%%%%%%%%%%%%%%%%%%%%%%%%%%%%%%%%%%%%%%%%

To resolve the above controversy, in this paper, we directly compute the quasiparticle spectral weight distribution by DMRG. Although $Z_{\mathrm {tot}}$ previously calculated by WSK can be indeed reproduced, we point out that it is not sufficient to conclude the Bloch-wave behavior of the doped hole. Rather, one has to further examine $Z_k$ and $Z_j$, denoting the probabilities of the ground state projecting onto a bare-hole Bloch state at momentum $k$ and site $j$, respectively. We determine $Z_k$ and $Z_j$, and show that the standard Bloch-wave behavior does break down on the one side of the aforementioned critical point, despite that $Z_{\mathrm {tot}}$ still remains finite and smooth. In particular, the length scale associated with the incoherence of the hole is identified from $Z_j$. We further derive an analytic formula serving as a direct probe of the
underlying mechanism responsible for the charge incoherence and modulation, which is also verified by the DMRG calculation.
%=====================================================Fig2=============================================================================
\begin{figure*}[tbp]
\begin{center}
\includegraphics[height=4in,width=6in]{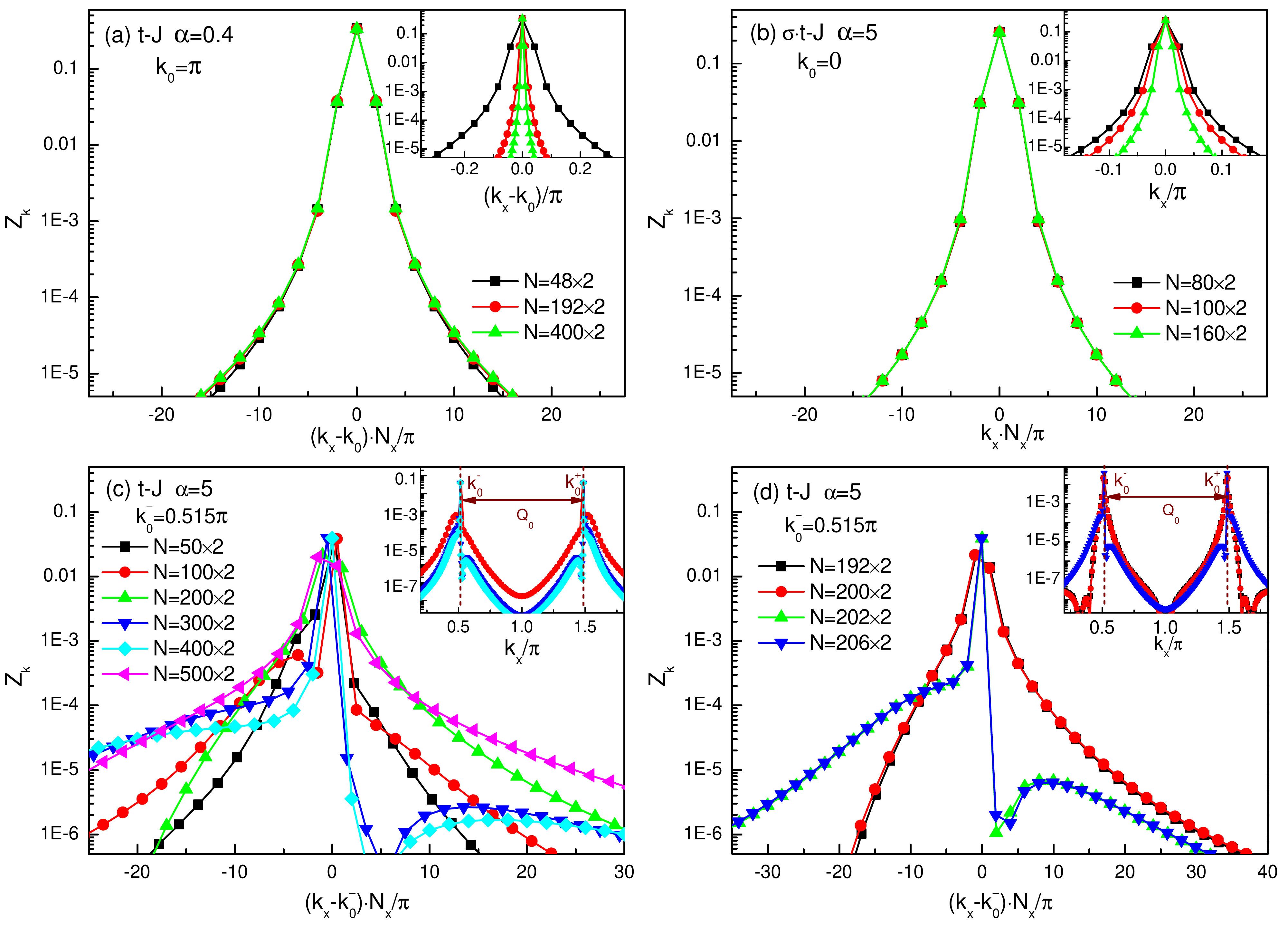}
\end{center}
\par
\renewcommand{\figurename}{Fig.}
\caption{(Color online). The quasiparticle spectral weight $Z_k$. Insets: the original $Z_k$'s in different models with various sample lengths. Main panels: Bloch-wave quantization under the OBC as characterized by the scaling law with the $k_x$-axis replaced by $(k_x-k_0)N_x/\pi$. (a) The $t$-$J$ case at $\alpha=0.4<\alpha_c$, a well-quantized Bloch wave with $k_0=\pi$; (b) The $\sigma$$\cdot$$t$-$J$ ladder at $\alpha=5$, a well-quantized Bloch wave with $k_0=0$; (c) and (d) The $t$-$J$ case at $\alpha=5>\alpha_c$: although $Z_{\mathrm{tot}}$ as the total sum well converges (cf. the inset of Fig. 1), the quantization in a finite-size sample breaks down due to strong phase shifts occurring even at small variations of the sample length, for example, $N_x=192$, $200$, $202$, and $206$ [cf. (d)]. Here $k_0$ is split into two $k^{\pm}_0$ separated by an incommensurate $Q_0$ [cf. the insets of (c) and (d)].
}
\label{Fig:Zk}
\end{figure*}
%================================================================================================================================

\emph{The model.---} We study the one-hole ground state based on the standard two-leg $t$-$J$ Hamiltonian composed of two one-dimensional chains  (each with the hopping integral $\alpha t$ and the superexchange coupling $\alpha J$), which are coupled together by the hopping $t$ and superexchange $J$ at each rung to form a two-leg ladder  \cite{ZZ2013,ZZ2014qp}. Here, the anisotropic parameter $\alpha\rightarrow 0$ in the strong rung limit, while two chains are decoupled at $\alpha \rightarrow \infty$. We focus on the model with $t/J=3$, which is the same as studied in Refs.~\onlinecite{ZZ2013,ZZ2014,ZZ2014qp,ZZ2014cm}.
%,Wang2015,WSK2015}.

For the one-hole-doped $t$-$J$ model, an exact expression of the partition function is given by \cite{Wu2008sign}
\begin{equation}\label{Zc}
\mathcal{Z}_{t\text{-}J}=\sum_{c}{\tau }_{c}\mathcal{W}[c]~,
\end{equation}
where the hole acquires a Berry-like phase \cite{Sheng1996} as
\begin{equation}\label{tauc}
\tau _c  = \left( { - 1} \right)^{N_h^ \downarrow  [c]}=\pm 1
\end{equation}
along a closed path $c$ (a brevity for multi-paths of the spins and the hole). Here $N_{h}^{\downarrow }[c]$ counts the total number of exchanges between the hole and down spins. The weight $\mathcal{W}[c]\ge 0$ is dependent on temperature ($1/\beta$), $t$, $J$, and $\alpha$ \cite{Wu2008sign}. The so-called $\sigma$$\cdot$$t$-$J$ model is introduced in Ref. \onlinecite{ZZ2013} by inserting a spin-dependent sign in the hopping term of the $t$-$J$ model, such that the one-hole partition function reduces to \cite{ZZ2013}
\begin{equation}\label{Zstj}
\mathcal{Z}_{\sigma\cdot t\text{-}J}=\sum_{c}\mathcal{W}[c]~,
\end{equation}
which is different from $\mathcal{Z}_{t\text{-}J}$ [Eq.~(\ref{Zc})] only by the absence of the Berry-like phase ${\tau }_{c}$, with the same $\mathcal{W}[c]$.

In the following, we shall study both models in a comparative way by using the DMRG algorithm \cite{ZZ2013,ZZ2014,ZZ2014qp,ZZ2014cm,WSK2015}. For these calculations,
we keep up to around $1800$ states, which controls the truncation error to be in the order of $10^{-10}$ and $10^{-6}$ for open and periodic systems, respectively.
For $Z_j$ calculations, we do more than 200 sweeps to obtain well converged results.

%%%%%%%%%%%%%%%%%%%%%%%%%%%%%%%Fig.3 %%%%%%%%%%%%%%%%%%%%%%%%%%%%%%%%%%%%%%%
\begin{figure*}[tbp]
\begin{center}
\includegraphics[height=3.2in,width=7.0in]{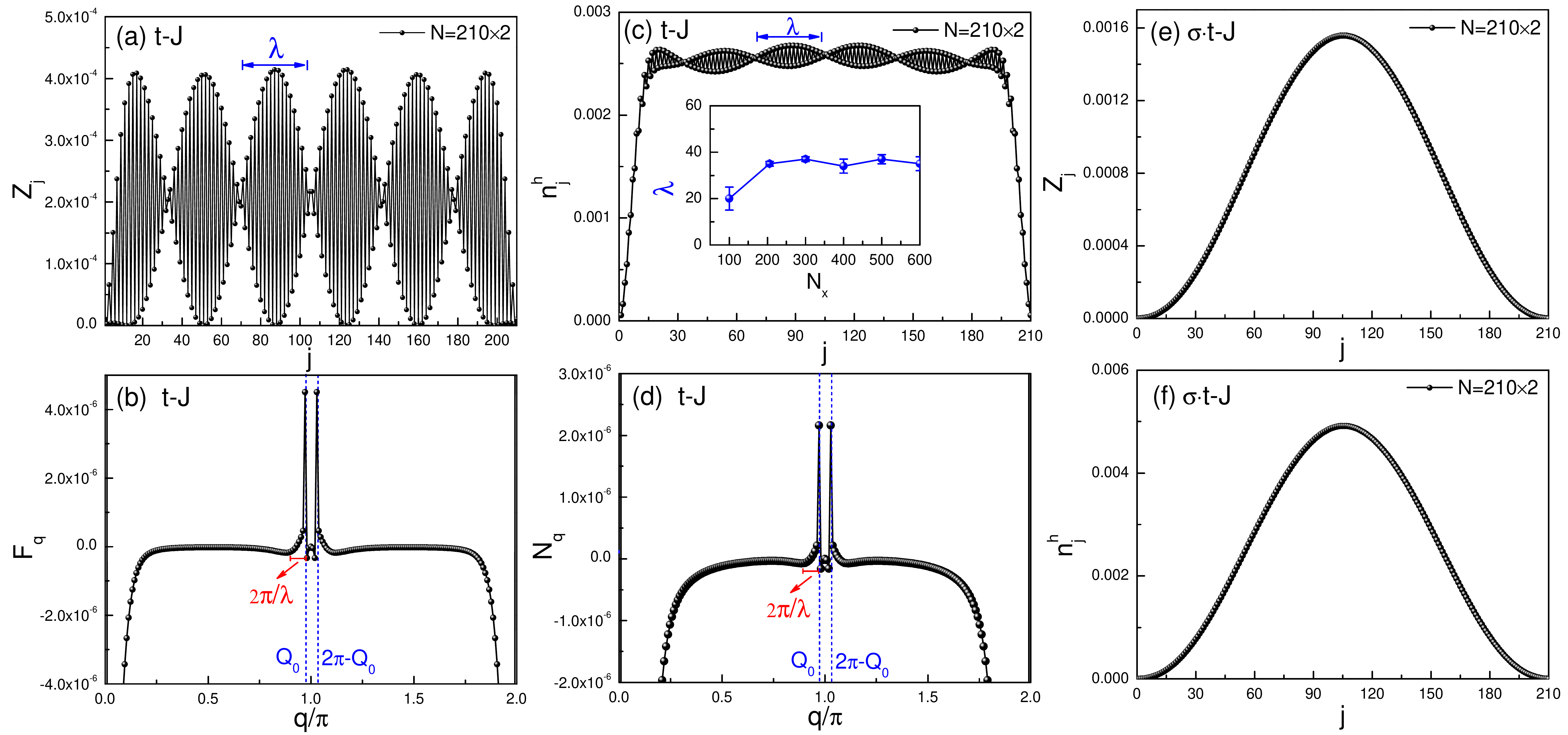}
\end{center}
\par
\renewcommand{\figurename}{Fig.}
\caption{(Color online). (a) $Z_j$ measures the probability of a bare hole at site $j$, which shows a fast oscillation modulated by a slower variation at a length scale of $\lambda$ ($\alpha=5>\alpha_c$); (b) The Fourier transformation of $Z_j$ reveals the characteristic wave vector $Q_0$ with a continuous spread $\sim 2\pi/\lambda$; (c) and (d) The corresponding hole density distribution $n_j^h$ and its Fourier transformation;  Inset of (c): The length scale $\lambda$ vs. $N_x$. Finally, smooth $Z_j$ [(e)] and $n_j^h$ [(f)] for the $\sigma\cdot$$t$-$J$ model at $\alpha=5$.  }
\label{Fig3}
\end{figure*}
%==================================================================================================================================

\emph{The quasiparticle spectral weight.---} The \emph{total} quasiparticle spectral weight is defined by
\begin{equation}\label{ztot}
Z_{\mathrm {tot}}\equiv \sum_k Z_k\equiv \sum_j Z_j~,
\end{equation}
where $Z_k\equiv \left |\langle {\bf k}|\Psi_G \rangle \right |^2$ or $Z_j\equiv \left |\langle j|\Psi_G \rangle \right |^2$ denotes the probability of the ground state $|\Psi_G \rangle$ in the bare hole Bloch state $|{\mathbf k}\rangle$ of momentum ${\mathbf k}$ or $|j\rangle$ at site $j$ of coordinate ${\mathbf r}_j$. Here $| {\mathbf k} \rangle \equiv \frac 1 {\sqrt {2N_x}} \sum_j \E^{i{\mathbf k}\cdot{\mathbf r}_j} |j\rangle $ and $|j\rangle\equiv \sqrt{2} c_j | \phi_0 \rangle$ (with a proper normalization factor included), where $| \phi_0 \rangle$ denotes the half-filling ground state. Note that ${\mathbf k}=(k_x, k_y)$ in general but we shall only focus on $k_y=0$  case in the considered regime of the two-leg ladder where the $k_y=\pi $ component of the $Z_k$ is exponentially small.

$Z_{\mathrm {tot}}$ computed by DMRG is shown in Fig.~\ref{Fig1}, which is in good agreement with the WSK's result \cite{WSK2015} for the $t$-$J$ case.
Note that $Z_{\mathrm {tot}}$ indeed remains a smooth function of $\alpha$ without exhibiting a singular behavior (though it decreases quickly) acrossing the critical $\alpha_c$. Here the $\alpha_c$ is marked by a vertical dashed line, which has been previously determined \cite{ZZ2014qp,ZZ2014cm} in terms of the ground state energy and the onset of a characteristic momentum $Q_0$ at $\alpha\geq\alpha_c$ [cf. Figs. 2(c) and 3(b)].
For comparison,  $Z_{\mathrm {tot}}$ for the $\sigma$$\cdot$$t$-$J$ ladder is also presented in Fig.~\ref{Fig1},
in which there is no critical point (with $Q_0=0$) throughout the whole $\alpha$ regime.

A finite $Z_{\mathrm {tot}}$ only means that $|\Psi_G \rangle$  has a finite probability remaining in a \emph{bare} hole state [cf. Eq.~(\ref{ztot})]. However, to determine whether the injected hole behaves like a Bloch wave or not, one needs to further inspect $Z_k$. Here $Z_k$ is found to be peaked at $k_0=\pi$ [or $k_0=0$] for the $t$-$J$ model at $\alpha<\alpha_c$ [or the $\sigma$$\cdot$$t$-$J$ model] as shown in Fig.~\ref{Fig:Zk}(a) [or (b)]. The data presented in the insets of Figs.~\ref{Fig:Zk}(a) and (b) can be well collapsed under a rescaling of $k_x$ by $(k_x-k_0)N_x$ in the main panels. They clearly indicate that the doped hole behaves like a coherent Bloch wave that is well \emph{quantized} in a finite size system [under an open boundary condition (OBC)]. In the large $N_x$ limit, the ground state possesses a single momentum  $k_0$, which satisfies the translation symmetry as expected.

At $\alpha>\alpha_c$, the momentum $k_0$ is split by $Q_0$ as $k_0^{+}-k_0^{-}=Q_0$ for the $t$-$J$ model. The emerging double-peak structure centered at $k_0^{\pm}$ is shown in the inset of Fig.~\ref{Fig:Zk}(c) at $\alpha=5>\alpha_c$. The wave quantization under the OBC is no longer valid here, as clearly illustrated in the main panel of Fig.~\ref{Fig:Zk}(c). Here many momenta (instead of two $k_0^{\pm}$) are involved in the large-$N_x$ case, which implies a breakdown of the translation symmetry. As a matter of fact, the distribution of momenta strongly scatter around $k_0^{\pm}$ even under small changes of sample sizes, for example, $N_x=192$, $200$, $202$, and $206$, as shown in Fig.~\ref{Fig:Zk}(d). It indicates that a large fluctuation may occur in the phase shift \cite{Anderson,Sheng1996} of the wave due to strong scattering between the hole and spin background, which scrambles the momentum quantization of the wave under the OBC. By contrast, $Z_{\mathrm{tot}}$ as the summation of $Z_k$ still converges quickly with the increase of $N_x$ [cf. the inset of Fig.~\ref{Fig1}].

To further verify the breakdown of the Bloch wave behavior observed above, one can examine the corresponding real space distribution $Z_j$. In Figs.~\ref{Fig3}(a) and (b), $Z_j$ and its Fourier transformation $F_q$ are presented, respectively, which exhibit a sharp spatial oscillation characterized by $Q_0$ [cf. Fig.~\ref{Fig3}(b)] at $\alpha=5$. Figure~\ref{Fig3}(a) further indicates another slower spatial modulation of a length scale $\lambda$, which corresponds to a continuous broadening around $Q_0$ in Fig.~\ref{Fig3}(b). It is consistent with the momentum smearing manifested in $Z_k$ around $k_0^{\pm}$ in Figs.~\ref{Fig:Zk}(c) and (d). Furthermore, the hole density distribution $n_j^h$ and its Fourier transformation $N_q$ are given in Figs.~\ref{Fig3}(c) and (d), which exhibit a charge modulation as well. In fact, by comparison it is easy to determine that the dominant contribution to the charge modulation in $n_j^h$ comes from $Z_j$, i.e., the bare hole component of $|\Psi_G \rangle$. The incoherent length scale $\lambda$ vs. $N_x$ is plotted in the inset of Fig.~\ref{Fig3}(c).
%----------------------------------------------------------------------------Fig.4-------------------------------------------------------------------------------------------------
\begin{figure}[t]
\begin{center}
\includegraphics[width=\columnwidth]{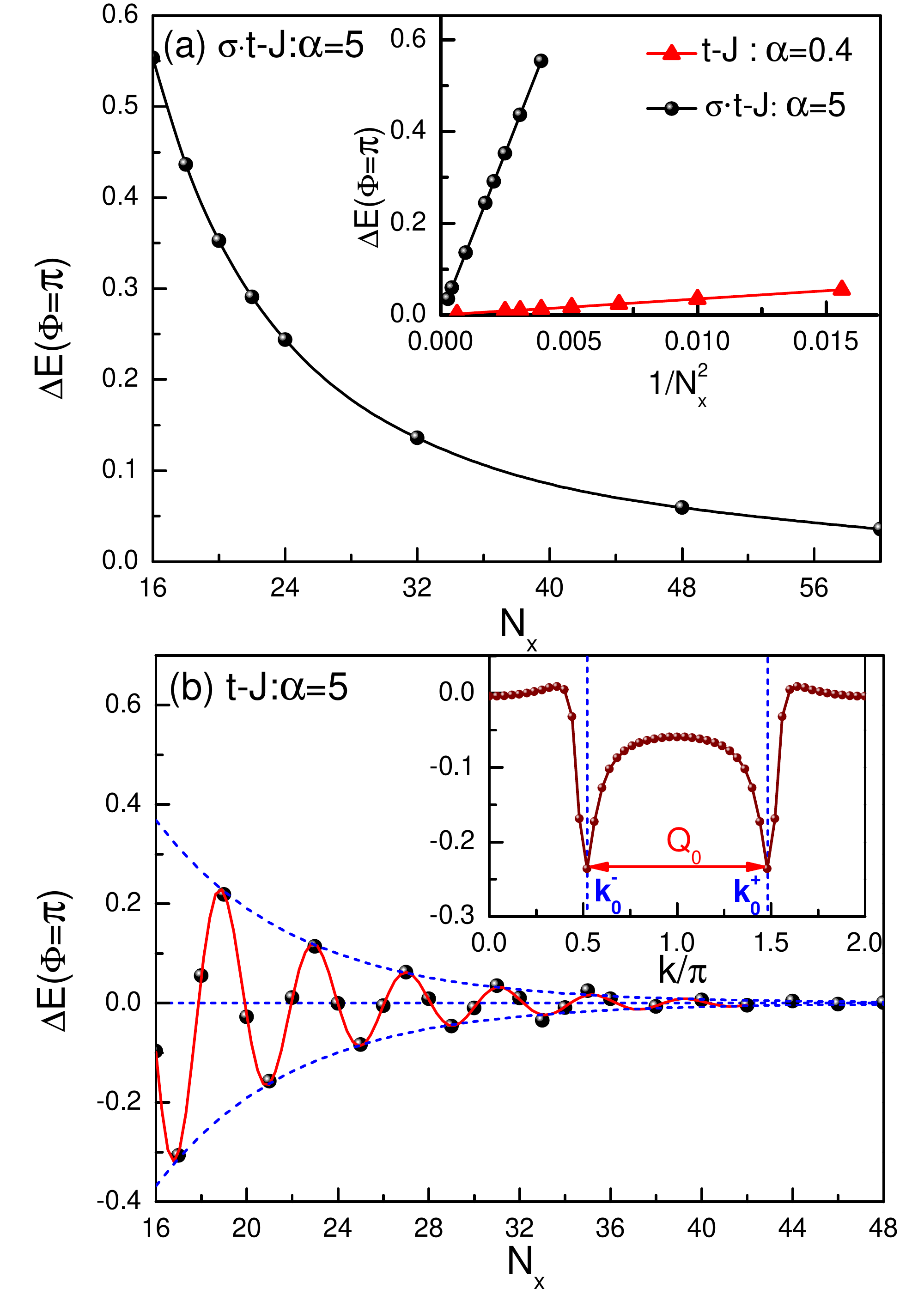}
\end{center}
\par
\renewcommand{\figurename}{Fig.}
\caption{(Color online). The energy change due to the charge response to an inserted  flux $\Phi=\pi$ into the ring geometry of the ladder (see text). (a) The typical Bloch wave behavior ($\propto 1/N_x^2$) for the $\sigma$$\cdot$$t$-$J$ case and the $t$-$J$ ladder at $\alpha<\alpha_c$ (the inset); (b) The non-Bloch-wave response at $\alpha>\alpha_c$ in the $t$-$J$ case can be well fitted by Eq.~(\ref{eq:deltaE1}), which directly relates the phase string effect as the underlying cause for charge incoherence and incommensurate momentum splitting (the inset) according to Eq.~(\ref{eq:deltaE}).}
\label{Fig:inch}
\end{figure}
%---------------------------------------------------------------------------------------------------------------------------------------------------------------------------------------

It is noted that $Z_k$ and $Z_j$ are determined by the single hole propagator, which may be formally expressed as \cite{Sheng1996,Wu2008sign} $G_h(i, j; E)\propto \sum_{c_{ij}}\tau_{c_{ij}} P(c_{ij})$ where $c_{ij}$ include all the paths of spins and the hole with the hole path connecting site $i$ and $j$, and the weight $P(c_{ij})>0$ \footnote{With $E$ less than the ground state energy $E_G^{\text{1-hole}}$}. According to Eq.~(\ref{tauc}), one may show \cite{ZZ2014cm} that the momentum structure and the charge modulation come from $\tau_{c_{ij}} \sim \E^{i\mathbf {k}_0\cdot [{\mathbf r}_i-{\mathbf r}_j]+i\delta_{ij}} $, in which $\mathbf {k}_0 \cdot [{\mathbf r}_i-{\mathbf r}_j]$ denotes an averaged $N_h^ \downarrow(c_{ij})$ and the phase shift $\delta_{ij}$ captures the rest of many-body fluctuations around $\mathbf {k}_0=\mathbf {k}_0^{\pm}$. The phase shift $\delta_{ij}$ is the source leading to the above breakdown of the Bloch wave behavior. As a matter of fact, by switching off $\tau_c$ in the $\sigma\cdot$$t$-$J$ model, all the modulations disappear in $Z_j$ and $n_j^h$ as indicated in Figs.~\ref{Fig3}(e) and (f).

\emph{Charge incoherence.---}  To probe the charge incoherence revealed by $Z_k$ and $Z_j$ at $\alpha>\alpha_c$, one may alternatively study the charge response to inserting a magnetic flux $\Phi$ into a ring of the ladder enclosed along the chain direction. Define the energy change $\Delta E_{G}^{\text{1-hole}} \equiv E_{G}^{\text{1-hole}} (\Phi=\pi) - E_{G}^{\text{1-hole}} (\Phi=0)$, with $\Phi=0$ corresponding to the periodic boundary condition (PBC) and $\Phi=\pi$ the anit-PBC for the hole \cite{ZZ2013,ZZ2014qp}. For a Bloch-wave behavior of the doped hole, one expects \cite{ZZ2013,ZZ2014qp} that $\Delta E_{G}^{\text{1-hole}} \propto1/N_x^2$. Indeed, as confirmed by DMRG, this is true for the $\sigma\cdot$$ t$-$J$ case [Fig. 4(a) and the inset] as well as the $t$-$J$ model at $\alpha<\alpha_c$ [the inset of Fig. 4(a), $N_x=\mathrm{even}$].

However, for the $t$-$J$ case at $\alpha>\alpha_c$, the charge incoherence is clearly manifested as shown in Fig.~\ref{Fig:inch}(b) at $\alpha=5$: $ \Delta E_{G}^{\text{1-hole}} $ oscillates strongly with $N_x$, which can be fitted by
\begin{equation}\label{eq:deltaE1}
\Delta E_{G}^{\text{1-hole}} (t\text{-}J) \propto \left(\E^{ik_0^+ N_x}+\E^{ik_0^- N_x}\right) g({N_x})~,
\end{equation}
where the incommensurate $k_0^{\pm}$ emerge as indicated in the inset of Fig.~\ref{Fig:inch}(b). Here the envelope function, $g(N_x)$, further gives rise to the broadening of the peaks $k_0^{\pm}$ as shown by its Fourier transformation in the inset of Fig.~\ref{Fig:inch}(b), which characterizes the incoherent scale of the charge. \footnote{Here one can fit $g\sim \E^{-\frac {N_x}{\xi}}$ with $\xi\sim 6.3$ as a characteristic incoherence (localization) length scale \cite{ZZ2013}. Note that $\xi$ and $\lambda/2$ ($\sim 16$), determined by the density distribution under OBC in Fig.~\ref{Fig3}(c), are related but not necessarily the same due to different ways of measurement. Further, the interference of the phase string under the PBC may be stronger at a relatively smaller $N_x$ because there are more channels for the destructive interference involving the hole circumventing the closed ring. The results under PBC are expected to be eventually in full agreement with those under OBC at sufficiently larger $N_x$, which is beyond our current computing capability. }

Analytically, a straightforward manipulation in terms of Eq.~(\ref{Zc}) gives rise to
\begin{align}
  \label{eq:deltaE}\nonumber
  \Delta E_{G}^{\text{1-hole}} (t\text{-}J) &= -\lim_{\beta\to\infty} \frac{1}{\beta} \ln \left( \frac{\mathcal{Z}_{t\text{-}J}(\Phi=\pi)}{\mathcal{Z}_{t\text{-}J}(\Phi=0)}  \right)\\
& =2 \sum_{c_1}  \tau_{c_1} \rho_{c_1}  + 2\sum_{c_3} \tau_{c_3} \rho_{c_3}+...  ~,
\end{align}
while, for the $\sigma$$\cdot$$t$-$J$ model, $\Delta E_{G}^{\text{1-hole}} (\sigma t\text{-}J)  =2 \sum_{c_1}   \rho_{c_1}  + 2\sum_{c_3}  \rho_{c_3}+... $ \footnote{By using the fact that each term is vanishingly small in the large $N_x$ limit, cf. Fig.~\ref{Fig:inch}(a).}. Here,
$\mathcal{Z}_{t\text{-}J}(\Phi=0)\equiv\sum_{\nu}\mathcal{Z}_{t\text{-}J}^{(\nu)}$ and $\mathcal{Z}_{t\text{-}J}(\Phi=\pi)\equiv\sum_{\nu}(-1)^{\nu}\mathcal{Z}_{t\text{-}J}^{(\nu)}$ with $\mathcal{Z}_{t\text{-}J}^{(\nu)}\equiv\sum_{c_{\nu}}{\tau }_{c_{\nu}}\mathcal{W}[c_{\nu}]$, where $\nu$ denotes the winding number counting how many times the hole circumvents the ring, and $\rho_{c_{\nu}}\equiv \lim_{\beta\to\infty} \mathcal{W}[c_{\nu}]/(\beta \mathcal{Z}_{t\text{-}J}^{(0)})>0$. Therefore, Eq.~(\ref{eq:deltaE1}) and Fig.~\ref{Fig:inch} provide a direct measurement of $\sum_{c_1}  \tau_{c_1} \rho_{c_1}$ in Eq.~(\ref{eq:deltaE}) at large $N_x$ (note that $\nu>1$ terms decay faster as $N_x$ increases), which indeed gives rise to the incommensurate $k_0^{\pm}$ and relates an incoherence scale with the phase shift fluctuation in $\tau_{c_1}$. These are indeed consistent with the picture previously obtained based on $Z_k$ and $Z_j$.

\emph{Conclusions.---} Doping into a gapped spin system is one of the simplest cases of doped Mott physics. Should the non-Bloch-wave behavior be fully validated for the one hole case, complemented by a strong pairing discovered \cite{ZZ2014} for the two hole case, an important understanding of the nature of strong correlation can be gained. Both DMRG simulations on the quasiparticle spectral weights, $Z_k$ and $Z_j$, and the charge response to inserting a flux have given rise to a consistent picture in this work. Namely, the momentum splitting and charge modulations found \cite{ZZ2013,ZZ2014qp,ZZ2014cm} at $\alpha>\alpha_c$ cannot be simply reduced to a standing wave description of two counter-propagating Bloch waves \cite{WSK2015}. Here, the hole loses its Bloch wave coherence intrinsically with involving a continuum of momenta, which indicates a spontaneous translational symmetry breaking. The microscopic origin due to $\tau_c$ in Eq.~(\ref{tauc}) (which is called the phase string effect \cite{Sheng1996} characterizing the long-range entanglement between the spins and the doped charge\cite{zaanen2011,Weng2011b}) has been thus established. This mechanism has also been recently studied  by a variational wave function approach \cite{Wang2015}, which can reproduce, e.g., $\alpha_c$ and $Q_0$, found by DMRG.

\begin{acknowledgements}
Useful discussions with R.-Q. He, H.-C. Jiang, Y. Qi, J. Zaanen are acknowledged. This work is supported by Natural Science Foundation of China (Grant No. 11534007), National Program for Basic Research of MOST of China (Grant No. 2015CB921000), and US National Science Foundation Grant  DMR-1408560.
\end{acknowledgements}

\end{document}